\documentclass[twocolumn]{aastex6}
\usepackage{newtxtext,newtxmath}

\usepackage[T1]{fontenc}
\usepackage{ae,aecompl}

\usepackage{graphicx}	
\usepackage{epstopdf}
\usepackage{amsmath}	
\usepackage{amssymb}	
\usepackage{hyperref}
\usepackage{longtable}

\begin{document}

\title{Ionized and molecular gas kinematics in a $z=1.4$ star-forming galaxy}\thanks{
Based on observations carried out with the IRAM Interferometer NOEMA. IRAM is supported by INSU/CNRS (France), MPG (Germany) and IGN (Spain). 
Based on observations carried out with the LBT. The LBT is an international collaboration among institutions in the United States, Italy and Germany. LBT Corporation partners are: LBT Beteiligungsgesellschaft, Germany, representing the Max-Planck Society, The Leibniz Institute for Astrophysics Potsdam, and Heidelberg University; The University of Arizona on behalf of the Arizona Board of Regents; Istituto Nazionale di Astrofisica, Italy; The Ohio State University, and The Research Corporation, on behalf of The University of Notre Dame, University of Minnesota and University of Virginia.}

\author{H.~\"Ubler\altaffilmark{1},
R.~Genzel\altaffilmark{1,}\altaffilmark{2},
L.~J.~Tacconi\altaffilmark{1}, 			
N.~M.~Förster~Schreiber\altaffilmark{1}, 	
R.~Neri\altaffilmark{3},				
A.~Contursi\altaffilmark{1},	
S.~Belli\altaffilmark{1}, 		
E.~J.~Nelson\altaffilmark{1}, 	
P.~Lang\altaffilmark{4}, 
T.~T.~Shimizu\altaffilmark{1},
R.~Davies\altaffilmark{1}, 
R.~Herrera-Camus\altaffilmark{1}, 
D.~Lutz\altaffilmark{1}, 		
P.~M.~Plewa\altaffilmark{1},	
S.~H.~Price\altaffilmark{1},	
K.~Schuster\altaffilmark{3},
A.~Sternberg\altaffilmark{5,}\altaffilmark{6},
K.~Tadaki\altaffilmark{7}, 
E.~Wisnioski\altaffilmark{8},	
S.~Wuyts\altaffilmark{9}
}

\altaffiltext{1}{Max-Planck-Institut f\"ur extraterrestrische Physik, Gie{\ss}enbachstr.\ 1, D-85748 Garching, Germany \href{mailto:hannah@mpe.mpg.de}{(hannah@mpe.mpg.de)}}
\altaffiltext{2}{Departments of Physics and Astronomy, University of California, Berkeley, CA 94720, USA}
\altaffiltext{3}{Institut de Radioastronomie Millimétrique, 300 rue de la Piscine, F-38406 Saint Martin d'Hères, France}
\altaffiltext{4}{Max-Planck-Institut f\"ur Astronomie, K\"onigstuhl 17, D-69117 Heidelberg, Germany}
\altaffiltext{5}{Raymond \& Beverly Sackler School of Physics \& Astronomy, Tel Aviv University, Ramat Aviv 69978, Israel}
\altaffiltext{6}{Center for Computational Astrophysics, Flatiron Institute, 162 Fifth Avenue, New York, NY 10010, USA}
\altaffiltext{7}{National Astronomical Observatory of Japan, 2-21-1 Osawa, Mitaka, Tokyo 181-8588, Japan}
\altaffiltext{8}{Research School of Astronomy \& Astrophysics, Australian National University, Canberra, ACT-2611, Australia}
\altaffiltext{9}{Department of Physics, University of Bath, Claverton Down, Bath, BA2 7AY, United Kingdom}

\begin{abstract}
We present deep observations of a $z=1.4$ massive, star-forming galaxy in molecular and ionized gas at comparable spatial resolution (CO 3-2, NOEMA; H$\alpha$, LBT). 
The kinematic tracers agree well, indicating that both gas phases are subject to the same gravitational potential and physical processes affecting the gas dynamics. 
We combine the one-dimensional velocity and velocity dispersion profiles in CO and H$\alpha$ to forward-model the galaxy in a Bayesian framework, combining a thick exponential disk, a bulge, and a dark matter halo. 
We determine the dynamical support due to baryons and dark matter, and find a dark matter fraction within one effective radius of $f_{\rm DM}(\leq$$R_{e})=0.18^{+0.06}_{-0.04}$. 
Our result strengthens the evidence for strong baryon-dominance on galactic scales of massive $z\sim1-3$ star-forming galaxies recently found based on ionized gas kinematics alone. 
\end{abstract}




\section{Introduction}\label{intro}

\begin{figure*}
	\centering
	\includegraphics[width=\textwidth]{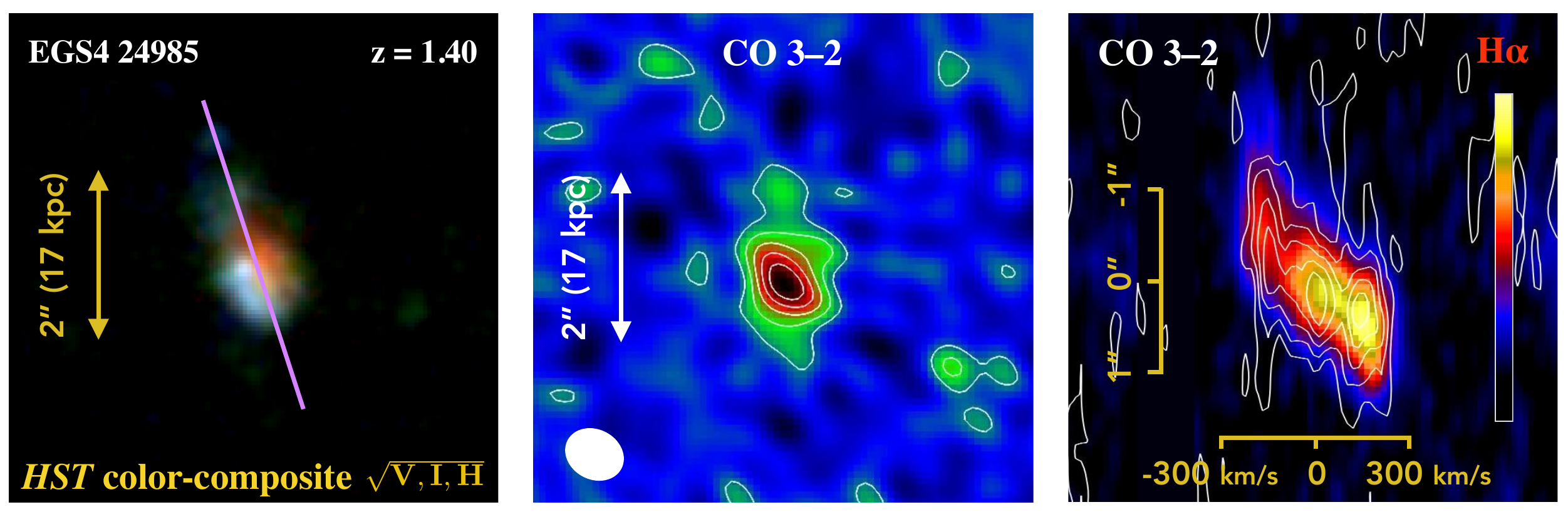}
    \caption{{\bf Left:} HST color-composite image of EGS4-24985. The magenta line shows the morphological position angle. {\bf Middle:} Uniformly weighted CO(3-2) image. The white ellipse shows the clean beam. {\bf Right:} H$\alpha$ (intensity color scale) and CO (white intensity contours) PV diagram.}
    \label{fig:fig1}
\end{figure*}

Our knowledge of the kinematics of star-forming galaxies (SFGs) at $z=1-3$ is dominated by large surveys targeting ionized gas emission \citep[e.g.][]{FS09, Kriek15, Wisnioski15, Stott16, Turner17}. There is strong evidence that the ionized gas kinematics of massive, high-redshift SFGs are dominated by ordered disk rotation, but a key question remains: how do the ionized gas kinematics, particularly the rotation curve and intrinsic velocity dispersion, compare to that of neutral or molecular gas, which dominate the gas mass budget?

Multi-phase, spatially-resolved data exist only for a handful of high-redshift SFGs, where the kinematics of the different gas phases are found to agree \citep[e.g.][]{Chen17}, or not (\citealp[e.g.][]{Swinbank11} {\it{vs.\ }}\citealp{Olivares16}). Yet, deeper data are generally needed for at least one of the gas phases in these studies to compare the kinematics in detail, and to disentangle the contributions from baryons and dark matter.
\cite{Genzel13} showed through deep integrations of a $z=1.5$ galaxy that its kinematics in H$\alpha$ and CO(3-2) agree. However, this galaxy is undergoing a minor merger and is therefore not optimally suited to kinematically analyze the galaxy's baryon {\it{vs.\ }}dark matter content. 

In this Letter, we analyze the H$\alpha$ and CO(3-2) kinematics of a massive SFG at $z=1.4$, EGS4-24985. We have obtained deep data, 21 and 45hrs on source, with the Large Binocular Telescope (LBT) and the NOrthern Extended Millimeter Array (NOEMA), making this an unprecedented data set of two important tracers of the gas kinematics in an SFG. We model the galaxy by combining a thick exponential disk, a bulge, and an NFW \citep{NFW96} halo, using Markov chain Monte Carlo (MCMC) sampling. We discuss correlations among the model parameters and constrain the galaxy's dark matter fraction within one effective radius ($R_{e}$). 
Throughout, we adopt a \cite{Chabrier03} initial mass function and a flat $\Lambda$CDM cosmology with $H_{0}=70$kms$^{-1}$Mpc$^{-1}$, $\Omega_{\Lambda}=0.7$, and $\Omega_{m}=0.3$.\\

\section{Data}\label{data}

\subsection{Physical Properties of EGS4-24985}\label{physprop}

EGS4-24985 (R.A.\ 14h19m26.66s, Dec.\ +52$^{\circ}51^{\prime}17.0\arcsec$) is a $z=1.4$ galaxy with a stellar mass of $M_{\star}=7.4\times10^{10}M_{\odot}$ and a star formation rate of SFR=$98.8M_{\odot}{\rm yr}^{-1}$ \citep[both derived following the techniques outlined by][]{WuytsS11a}, placing it in the upper half of the main sequence at this redshift \citep{Whitaker14}. 
The $V-,I-,H-$band ACS and WFC3 images reveal strong spatial color variations, indicative of a mixture of stellar populations, or varying dust obscuration that potentially hides a central mass concentration (Figure~\ref{fig:fig1}, left).

The morphological position angle PA$_{H}=18^{\circ}$, minor-to-major axis ratio $q_{H}=0.60$, $R_{e, H}=0\farcs52=4.4$kpc, and Sérsic index $n_{S,H}=0.74$ are constrained from {\sc galfit} \citep{Peng10} Sérsic models based on the 3D-HST team \citep{Skelton14} version of CANDELS $H$-band (F160W) imaging \citep{Grogin11, Koekemoer11}, presented by \cite{vdWel12}. Assuming a ratio of scale height to scale length of $q_{0}=0.2$, typical for SFGs at this redshift \citep[e.g.][]{vdWel14a}, the estimated inclination is $i=55^{\circ}$. There is a systematic change in $q$ as derived from other filters, $q=0.54-0.66$ from F125W ($J$-band) to F814W ($I$-band).

Assuming a bulge-to-disk decomposition with $n_{S,\rm disk}=1$, $n_{S,\rm bulge}=4$, we infer the bulge-to-total fraction from the stellar mass map to be $B/T=0.13\pm0.15$ \citep{Lang14}.

\begin{figure*}
	\centering
	\includegraphics[width=0.45\textwidth]{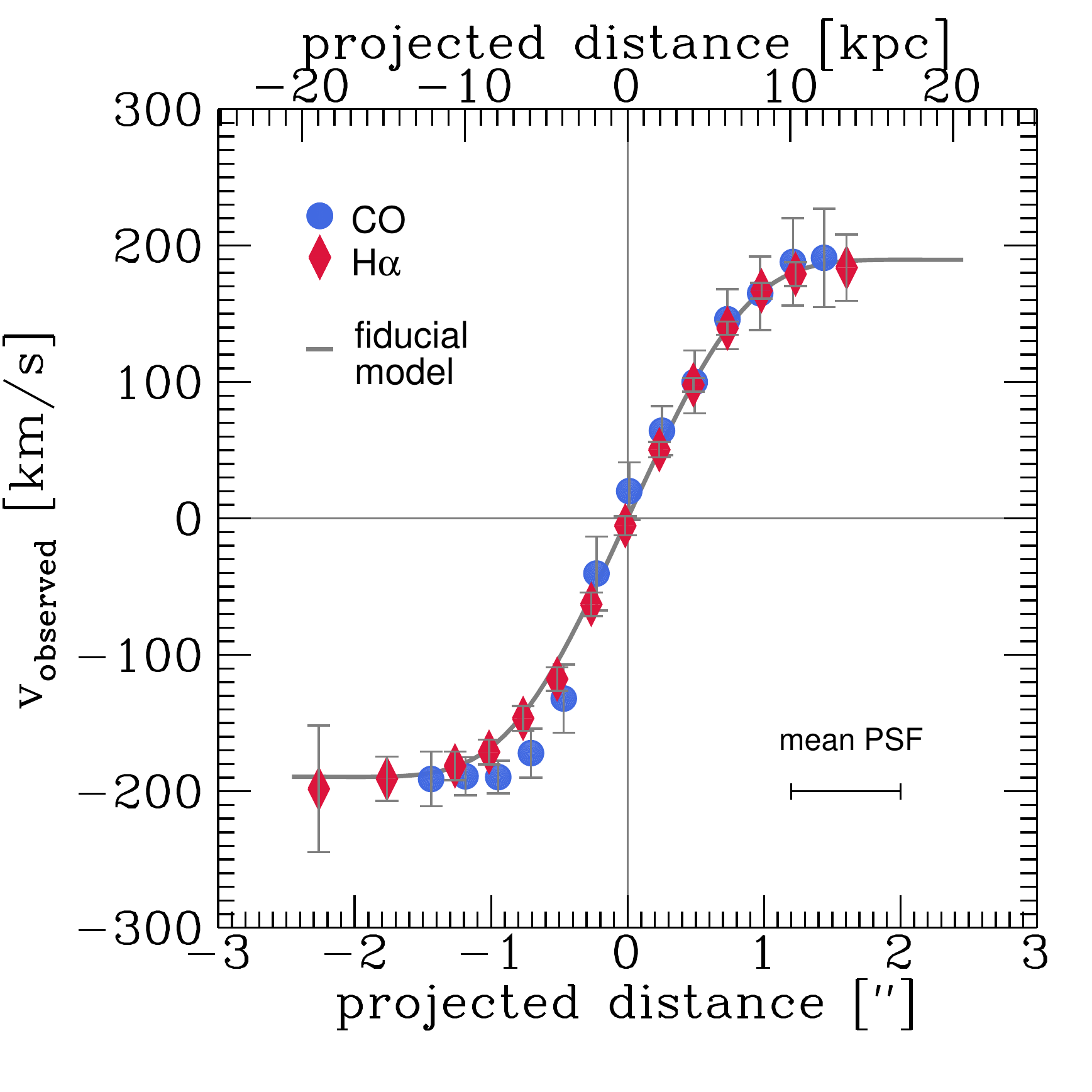}
	\hspace{10mm}
	\includegraphics[width=0.45\textwidth]{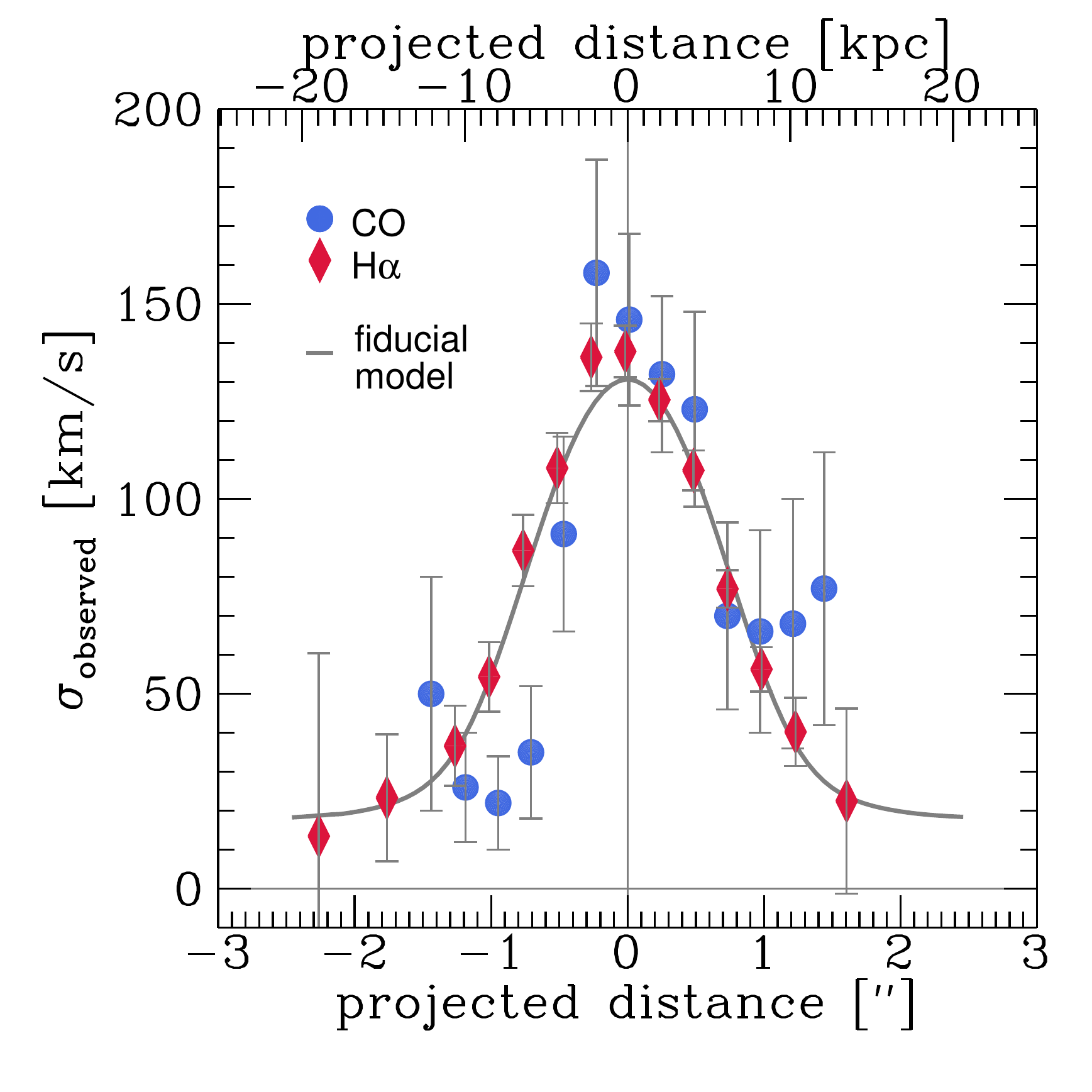}
    \caption{One-dimensional velocity (left) and velocity dispersion (right) profiles along the kinematic major axis in CO(3-2) (blue circles) and H$\alpha$ (red diamonds). Due to the spatial resolution of the observations, neighboring data points are not independent. The projected distance increases from NE to SW of the kinematic center of the galaxy. The extractions from the two tracers agree. In grey we show our fiducial model (see \S\ref{fiducial}).}
    \label{fig:fig2}
\end{figure*}

\subsection{CO observations with NOEMA}\label{co}

To explore the kinematics of the cold gas, we observed the CO(3-2) line 
with the IRAM interferometer NOEMA. At the redshift of the source, 
the CO(3-2) line (rest frequency 345.796GHz) is shifted into the 2mm 
band. We observed EGS4-24985 in the D (compact) and A (extended) 
configurations with 7 or 8 antennas between November 2016 and 
April 2017. The total equivalent 8-antenna on-source integration time 
was 45hrs in the D+A configuration, with a resolution of $0\farcs6-1\farcs0$. 
Weather conditions during the observing periods were excellent, with 
typical system temperatures of $\sim$150K. The WiDEX spectral correlator 
provided 4GHz of bandwidth per polarization with a fixed channel 
spacing of 2MHz. For phase and amplitude calibration, every 20 minutes 
we alternated source observations with observations of a bright quasar 
within $15^{\circ}$ of the source. The absolute flux calibration was done through bootstrapping from 
observations of LkHA-101 and MWC-349 (0.36Jy and 1.45Jy at 144GHz), 
 resulting in a continuum flux of $70\mu$Jy.

The data were calibrated using the CLIC package of the IRAM GILDAS\footnote{
\url{http://www.iram.fr/IRAMFR/GILDAS}} 
software environment, and imaged and analyzed with the MAPPING 
routines in GILDAS. We applied a uniform weighting scheme to create the 
data cube, and then subtracted the 2mm continuum emission using channels 
free of line emission. The final cube was CLEANED with the CLARK 
version of CLEAN implemented in GILDAS, and reconstructed with a 
$0\farcs67\times0\farcs55$ (PA=62$^{\circ}$) clean beam (Figure~\ref{fig:fig1}, middle), to a spectral resolution of 19kms$^{-1}$ with an rms noise of 0.2mJy~channel$^{-1}$synthesized beam$^{-1}$.

The molecular gas mass as measured from the CO(3-2) flux
and using the $\alpha$(CO) conversion function by \cite{Genzel15} is $M_{\rm mol}=6.9\times10^{10}M_{\odot}$. 
With a gas-to-baryonic mass fraction of $M_{\rm mol}/M_{\rm bar}=0.48$, the galaxy is typical when compared 
to larger samples at the same redshift \citep{Tacconi17}.
The CO distribution has an approximate extent of $R_{e, {\rm CO}}\approx0\farcs26$ (measured from an exponential disk fit in the UV plane). 
The CO position-velocity (PV) diagram is shown in Figure~\ref{fig:fig1} (right, white contours).

\subsection{H$\alpha$ observations with LUCI at LBT}\label{ha}

We obtained seeing-limited near-infrared spectroscopy of EGS4-24985 using 
the LUCI1 and LUCI2 spectrographs in binocular mode \citep{Buschkamp12}, mounted 
at the Bent Gregorian focus of the two 8.4m mirrors of the LBT \citep{Hill06}. 
The observations were carried out over five
nights in March 2017, in clear weather or thin clouds, with seeing 
$0\farcs6-1\farcs0$. We used a pixel scale of $0\farcs25$, the 210 grating
in $H-$band and a slit width of $1\farcs0$, yielding a spectral resolution of $R\sim3000$. We adopted a two-point dithering pattern and an exposure time
of 5min per frame, for a total on-source time of 21hrs (summed from both spectrographs). 
To facilitate acquisition, we used a multi-object mask and chose PA$=20^{\circ}$ 
to align the slit to the major axis of the galaxy (\S\ref{physprop}). The
data were reduced using the \texttt{flame} pipeline \citep{Belli17b}, which
outputs a rectified, sky-subtracted, wavelength-calibrated two-dimensional
(2D) spectrum. 
The corresponding H$\alpha$ PV diagram is shown in Figure~\ref{fig:fig1} (right, color scale).

\begin{table*}
\centering
 \caption{Results from our fiducial model and additional setups. We first list the model priors (`G(x, y)': Gaussian(center x, width y); `F[x; y]': flat prior in range [x; y]; `f: x': fixed to x) and then the medians with 1$\sigma$ confidence ranges of the marginalized probability distributions from the MCMC sampling.}
 \label{tab:tab1}
 \begin{tabular}{l|ccccccc}
 	model & {\bf fiducial}& && low $c$ & high $c$ & free halo, $B/T$ & 2 fixed disks \\
		& {\bf H$\alpha$+CO} & only H$\alpha$ & only CO & H$\alpha$+CO & H$\alpha$+CO & H$\alpha$+CO & H$\alpha$+CO \\
	\tableline	
	$M_{\rm bar}$ [$10^{11}M_{\odot}$] & G($1.4; 0.7$)&G($1.4; 0.7$) &G($1.4; 0.7$)& G($1.4; 0.7$) & G($1.4; 0.7$) & G($1.4; 0.7$) & f: 0.74 + 0.69 \\
		&  $1.1^{+0.5}_{-0.4}$ & $1.1^{+0.5}_{-0.3}$ & $1.2^{+0.5}_{-0.4}$& $1.2^{+0.4}_{-0.3}$ & $1.2^{+0.5}_{-0.4}$ & $1.0^{+0.5}_{-0.4}$ & --  \\
	$R_{e}$ [$\arcsec$] &G($0.52; 0.10$)& G($0.52; 0.10$) &G($0.52; 0.10$)& G($0.52; 0.10$) & G($0.52; 0.10$) & G($0.52; 0.10$) & f: 0.52; 0.26 \\
		& $0.49^{+0.10}_{-0.09}$ & $0.51^{+0.10}_{-0.10}$ & $0.50^{+0.10}_{-0.09}$& $0.53^{+0.09}_{-0.08}$ & $0.48^{+0.11}_{-0.10}$ & $0.50^{+0.10}_{-0.10}$ & --  \\
	$B/T$ & G($0.20; 0.15$)&G($0.20; 0.15$)&G($0.20; 0.15$) & G($0.20; 0.15$) & G($0.20; 0.15$) & F[0;1] & --  \\
		& $0.27^{+0.10}_{-0.09}$ & $0.28^{+0.10}_{-0.09}$ & $0.22^{+0.12}_{-0.11}$& $0.23^{+0.08}_{-0.07}$ & $0.29^{+0.11}_{-0.10}$ & $0.43^{+0.28}_{-0.18}$ & --  \\
	$i$ [$^{\circ}$] & G($55; 10$)&G($55; 10$) &G($55; 10$)& G($55; 10$) & G($55; 10$) & G($55; 10$) & f: 55 \\
		& $44^{+8}_{-6}$ & $44^{+9}_{-7}$ & $47^{+9}_{-7}$& $50^{+8}_{-7}$ & $39^{+9}_{-6}$ & $40^{+11}_{-9}$ & --  \\
	$\sigma_{0}$ [kms$^{-1}$] &G($30; 10$)& G($30; 10$)&G($30; 10$) & G($30; 10$) & G($30; 10$) & G($30; 10$) & F[5;100] \\
		& $17^{+5}_{-6}$ & $21^{+5}_{-6}$ & $19^{+7}_{-7}$& $16^{+5}_{-6}$ & $18^{+5}_{-6}$ & $17^{+5}_{-6}$ & $11^{+7}_{-4}$ \\
	$M_{\rm halo}$ [$10^{12}M_{\odot}$] &G($4.2; 2.0$)& G($4.2; 2.0$) &G($4.2; 2.0$)& G($4.2; 2.0$) & G($4.2; 2.0$) & F[0.001;100] & F[0.01;100] \\
		& $3.5^{+1.9}_{-1.7}$& $3.5^{+1.9}_{-1.7}$ & $3.9^{+1.9}_{-1.8}$ & $4.4^{+1.9}_{-1.9}$ & $2.2^{+1.9}_{-1.3}$ & $7.2^{+21}_{-5.1}$ & $0.015^{+0.011}_{-0.004}$ \\
	$c$ & f: 4.4 & f: 4.4 & f: 4.4 & f: 2 & f: 8 & f: 4.4 & F[1;10] \\
		& -- & -- & -- & --  & --  & -- & $1.3^{+0.5}_{-0.2}$ \\
	\tableline	
	inferred $f_{\rm DM}(\leq$$R_e$) & $0.18^{+0.06}_{-0.04}$ & $0.19^{+0.06}_{-0.05}$ & $0.20^{+0.08}_{-0.06}$ & $0.11^{+0.04}_{-0.03}$ & $0.25^{+0.07}_{-0.06}$ & $0.22^{+0.08}_{-0.06}$ & $0.008^{+0.002}_{-0.002}$ \\
 \end{tabular}
\end{table*} 

\subsection{One-dimensional kinematic profiles}\label{profiles}

To create the one-dimensional (1D) velocity and dispersion profiles in CO, we proceed as described by \cite{Genzel17}: we first fit a Gaussian profile to the CO line emission in each spaxel of the data cube, smoothed over three spaxels to ensure sufficient $S/N$ in the outer parts of the galaxy. Accounting for the galaxy's systemic velocity, this results in the 2D velocity map. From this we determine PA$_{\rm kin}=23^{\circ}$ as the axis with the steepest velocity gradient. It agrees with PA$_{H}$, and with the PA of the H$\alpha$ slit observations (\S\S\ref{physprop}, \ref{ha}).
The CO 1D velocity and dispersion profiles are then constructed from $0\farcs75$ diameter apertures (as a compromise between the CO data resolution and the seeing-limited H$\alpha$ data) with the center spaced by $0\farcs24$ along PA$_{\rm kin}$.

To create the 1D profiles in H$\alpha$, we extract spectra in overlapping bins of two to four spatial pixels and fit a Gaussian profile to the H$\alpha$ line emission. The choice of the number of spatial pixels used for the extraction of individual data points does not substantially affect the extracted values, but allows for increased $S/N$ in the outer disk regions.

We trace H$\alpha$ out to 19kpc (NE, $\sim4.4R_{e,H}$) and 13kpc (SW, $\sim3.1R_{e,H}$), and CO out to 12kpc ($\sim2.8R_{e,H}$). These physical radii at $z\sim1.4$ are equivalent to probing the rotation curve out to 23-35kpc for a galaxy of this stellar mass at $z\sim0$ \citep{vdWel14a}.
Figure~\ref{fig:fig2} shows the 1D velocity and dispersion profiles in CO and H$\alpha$ along PA$_{\rm kin}$ in observed space. The uncertainties are derived from the Gaussian fits described above where noise has been taken into account. The two tracers agree, indicating that they trace the same mass distribution, most reliably in the outer disk where beam-smearing effects become less important.

The galaxy's intrinsic velocity dispersion, $\sim$15-30kms$^{-1}$, is at the lower end of typical values of SFGs at this redshift ($\sim$45kms$^{-1}$; \citealp{Wisnioski15}; see also \citealp{DiTeodoro16}). This is evident from the outer regions of the 1D profile, where, under the assumption of constant intrinsic velocity dispersion, the effect of beam-smearing on the measured dispersion is low. Therefore, in the case of EGS4-24985, the correction for pressure support from the turbulent gas motions to the circular velocity is small ($\sim$8kms$^{-1}$ at $2\farcs3$), and thus does not lead to a significant drop in the observed outer rotation curve. 
Considering the limitations of the instrumental spectral resolution, the recovered dispersion values represent upper limits.\\

\section{Modelling}\label{modelling}

Since the 1D kinematic profiles of ionized and molecular gas agree within their uncertainties, it is justified to combine them to improve constraints on our model parameters.
We have also separately analyzed the H$\alpha$ and CO data and found agreement of the results within the uncertainties (Table~\ref{tab:tab1}).

The kinematic modelling of our galaxy follows the methodology described by \cite{WuytsS16} and \cite{Genzel17}.
We build a mass model consisting of a thick exponential disk ($n_{S}=1$, $q_{0}=0.2$) 
a bulge ($n_{S}=4$, $q_{0}=1$, $R_{e}=1$kpc), and an NFW halo. We fit the mass model simultaneously to the 1D velocity and dispersion profiles of H$\alpha$+CO along PA$_{\rm kin}$. For the baryonic mass distribution, we account for a finite flattening following \cite{Noordermeer08}. Our choice of an $n_{S}=1$ disk plus bulge is motivated by the bulge-to-disk decomposition and the likely high dust obscuration in the center of the galaxy.

The modelling uses an updated version of {\sc dysmal} \citep{Cresci09, Davies11, WuytsS16}. This code accounts for spectral and spatial beam-smearing, and incorporates the effects of pressure support on the circular velocity from the turbulent gas motions of the kinematic tracer, as described by \cite{Burkert10} and \cite{WuytsS16} \citep[see also][for a detailed discussion]{Dalcanton10}. 
The most important update to {\sc dysmal} consists of the implementation of an MCMC sampling procedure using the {\sc emcee} package \citep{FM13}. 
A full description of the updated code will be presented by Shimizu et al.\ ({\it in prep.}).

\begin{figure*}
	\centering
	\includegraphics[width=1.01\textwidth]{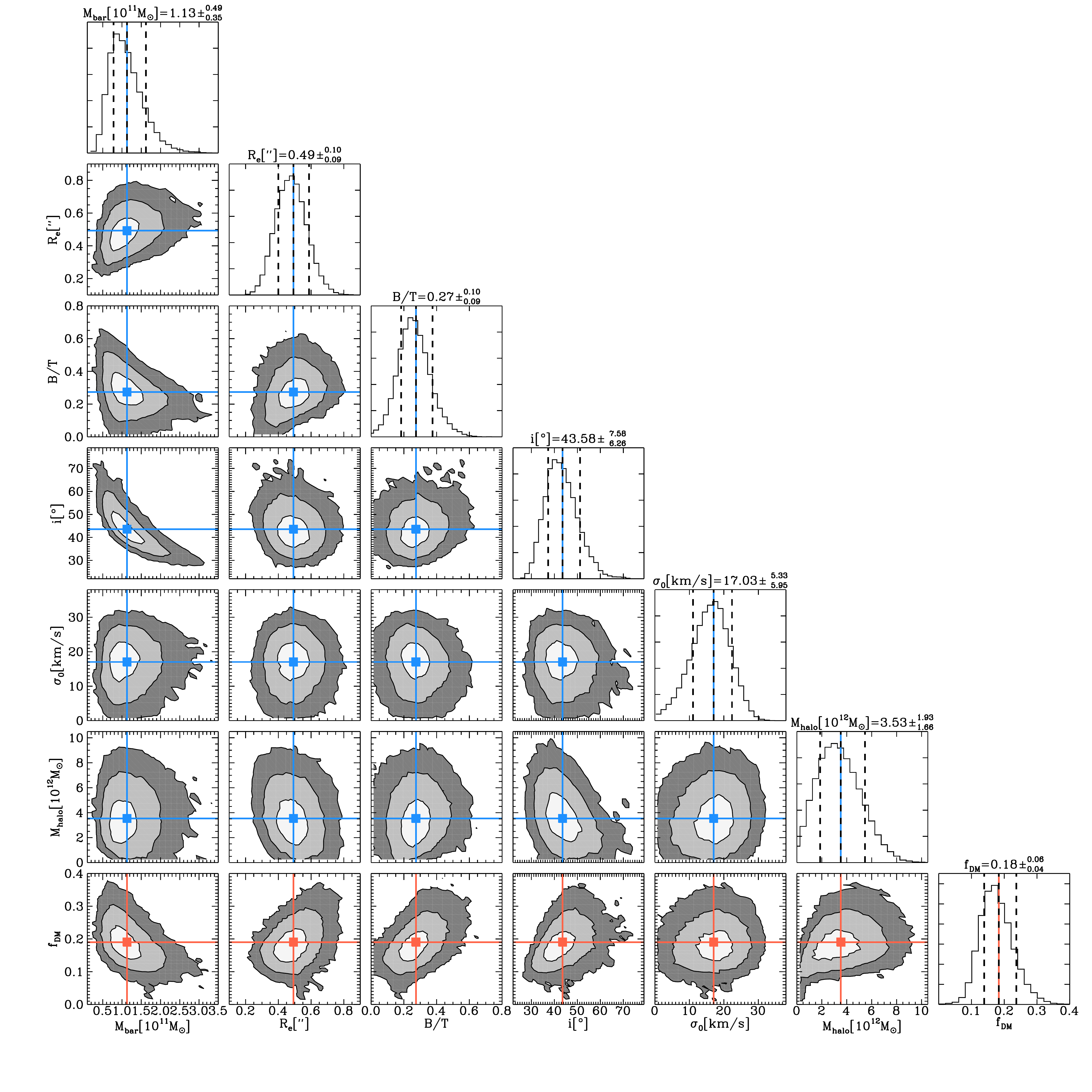}
    \caption{MCMC sampling of the joint posterior probability distribution of the fiducial model parameters, $M_{\rm bar}$, $R_e$, $B/T$, $i$, $\sigma_0$, and $M_{\rm halo}$ (top rows) from a combined fitting to the H$\alpha$+CO data. The median values and 1$\sigma$ confidence ranges of the marginalized distributions are indicated by the dashed vertical lines in the 1D histograms, and given on top of each histogram. The median values are also shown as blue squares on top of the 2D histograms. All of the median values lie close to the modes of the 2D histograms. The contours show the 1$\sigma$, 2$\sigma$, 3$\sigma$ confidence levels of the 2D distributions. 
    The bottom row histograms show $f_{\rm DM}(\leq$$R_{e})$, calculated from the intrinsic models. Median values are indicated in red. For the sampled parameter space, dark matter is sub-dominant within $R_e$.}
    \label{fig:fig3}
\end{figure*}

Free parameters in our modelling are $M_{\rm bar}$, $R_{e}$, $B/T$, $i$, $\sigma_{0}$, and the NFW halo mass $M_{\rm halo}$. 
We choose the prior halo mass to be typical for the redshift and stellar mass of our galaxy \citep{Moster13}. The concentration parameter is fixed to a value typical for this halo mass and redshift, $c=4.4$ \citep{Dutton14}. We verify that the typical concentration parameters for the range of halo masses derived from the MCMC sampling are broadly consistent with this value ($\Delta c\sim0.2$ for the 1$\sigma$ distribution of sampled halo masses). 
We explore setups with lower/higher concentrations ($c=2; 8$), and consequently find lower/higher $f_{\rm DM}(\leq$$R_{e})$ and higher/lower $M_{\rm halo}$, consistent with our main results (Table~\ref{tab:tab1}).
We do not consider adiabatic contraction since its net effect at high redshifts is not well constrained \citep[e.g.][]{Duffy10}.

In calculating the model likelihood, we assume Gaussian measurement noise. 
For the purpose of parameter inference, we choose Gaussian priors for all model parameters that reflect our prior state of knowledge about their values and uncertainties (Table~\ref{tab:tab1}).
As discussed in \S\ref{physprop}, $q$ and $R_{e}$ are independently constrained through {\sc galfit} models. The adopted uncertainties of $\sigma_{i}=10^{\circ}$ and $\sigma_{R_{e}}=0\farcs10$ are conservative estimates \citep[see][]{vdWel12}. Through our choice of narrow Gaussian priors for these parameters, we translate their uncertainties directly into the modelling.
We choose $B/T=0.2$ with $\sigma_{B/T}=0.15$ to account for a possible bulge hidden by dust extinction. 
For $M_{\rm bar}$ and $M_{\rm halo}$, we adopt uncertainties of $\sim50\%$. For $\sigma_0$, our estimate is roughly based on the outer values of the dispersion profile.
If we adopt flat priors for $M_{\rm halo}$ and $B/T$, we find consistent results. 
We also explored a model with fixed stellar and gaseous exponential disks, no bulge, free $\sigma_{0}$, $M_{\rm halo}$, and $c$, leading to a central dark matter fraction of $<1\%$ (Table~\ref{tab:tab1}).

For our fiducial model, we set up the MCMC sampling of the posterior probability function of the parameters with 180 walkers, a burn-in phase of 500 steps, and a running phase of 2000 steps. The length of the burn-in was designed to ensure convergence of the chains, while the length of the final run was designed to be >10 times the maximum autocorrelation time of the individual parameters. The acceptance fraction of the final run was 0.35.\\

\section{Results}\label{results}

\subsection{Parameter correlations and fiducial model}\label{fiducial}

The MCMC sampling of the joint posterior probability distributions of the model parameters is visualized in the top rows of Figure~\ref{fig:fig3}. 
The median values and 1$\sigma$ confidence ranges of the marginalized distributions are indicated by the dashed vertical lines in the 1D histograms (see also second column in Table~\ref{tab:tab1}).

For the 2D marginalized distributions, contours show the 1$\sigma$, 2$\sigma$, and 3$\sigma$ confidence levels. 
The strongest correlation is between inclination and $M_{\rm bar}$. This is expected, since any inclination correction to the observed rotation velocity directly affects the inferred dynamical mass. This is also reflected to a smaller extent in the correlation between inclination and $M_{\rm halo}$.

Since the posterior distribution is well behaved, we choose our fiducial model to be represented by the median values of the individual marginalized distributions, with uncertainties represented by the 1$\sigma$ confidence ranges. The median values are also shown as blue squares in the 2D histograms in Figure~\ref{fig:fig3}. Every median lies close to the mode of the posterior distribution in projection, indicating that they lie in the most likely parameter space.

The 1D profiles of velocity and dispersion corresponding to the fiducial model in observed space are shown as grey lines in Figure~\ref{fig:fig2}.

\subsection{Central dark matter fraction}

\begin{figure*}
	\centering
	\includegraphics[width=0.45\textwidth]{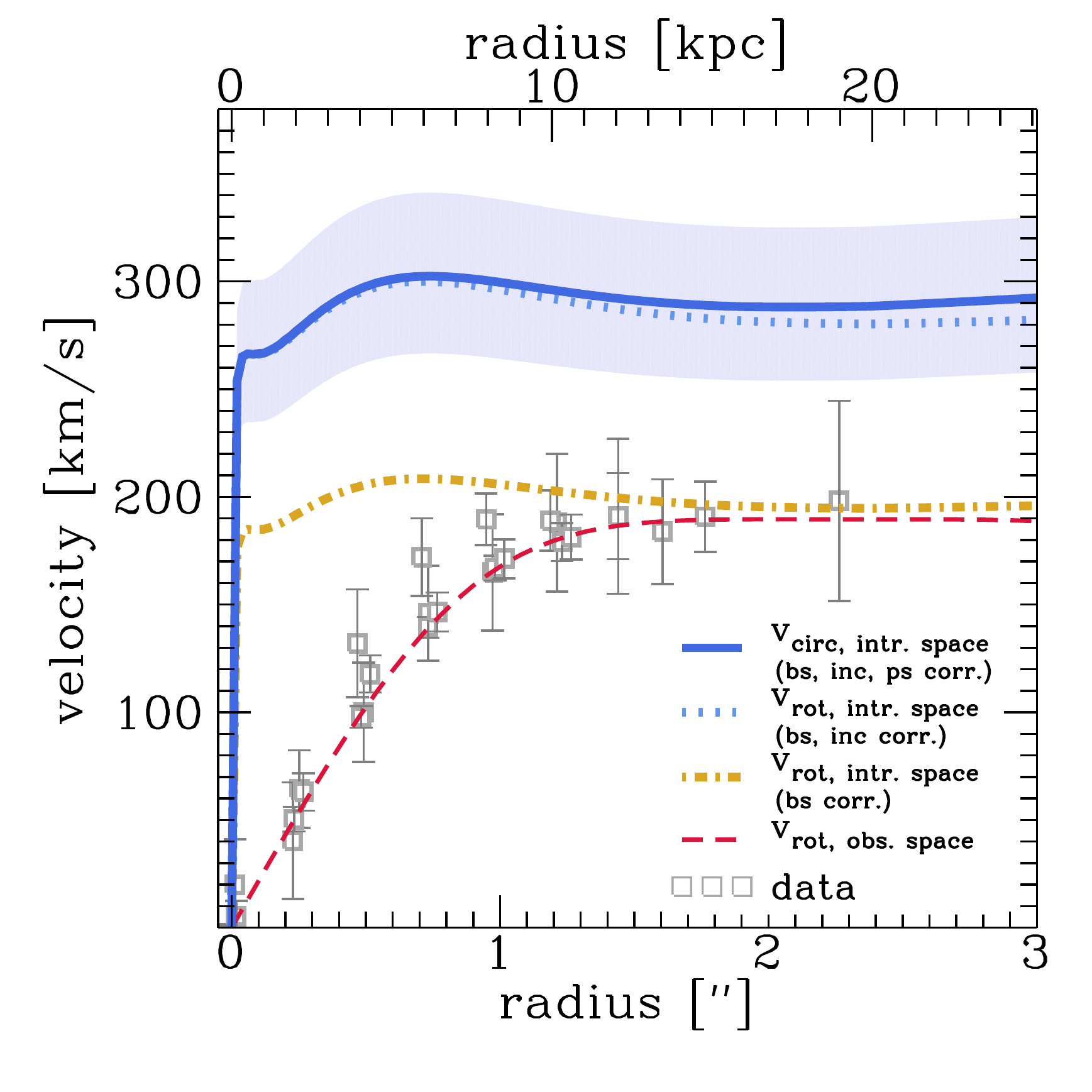}
	\hspace{10mm}
	\includegraphics[width=0.45\textwidth]{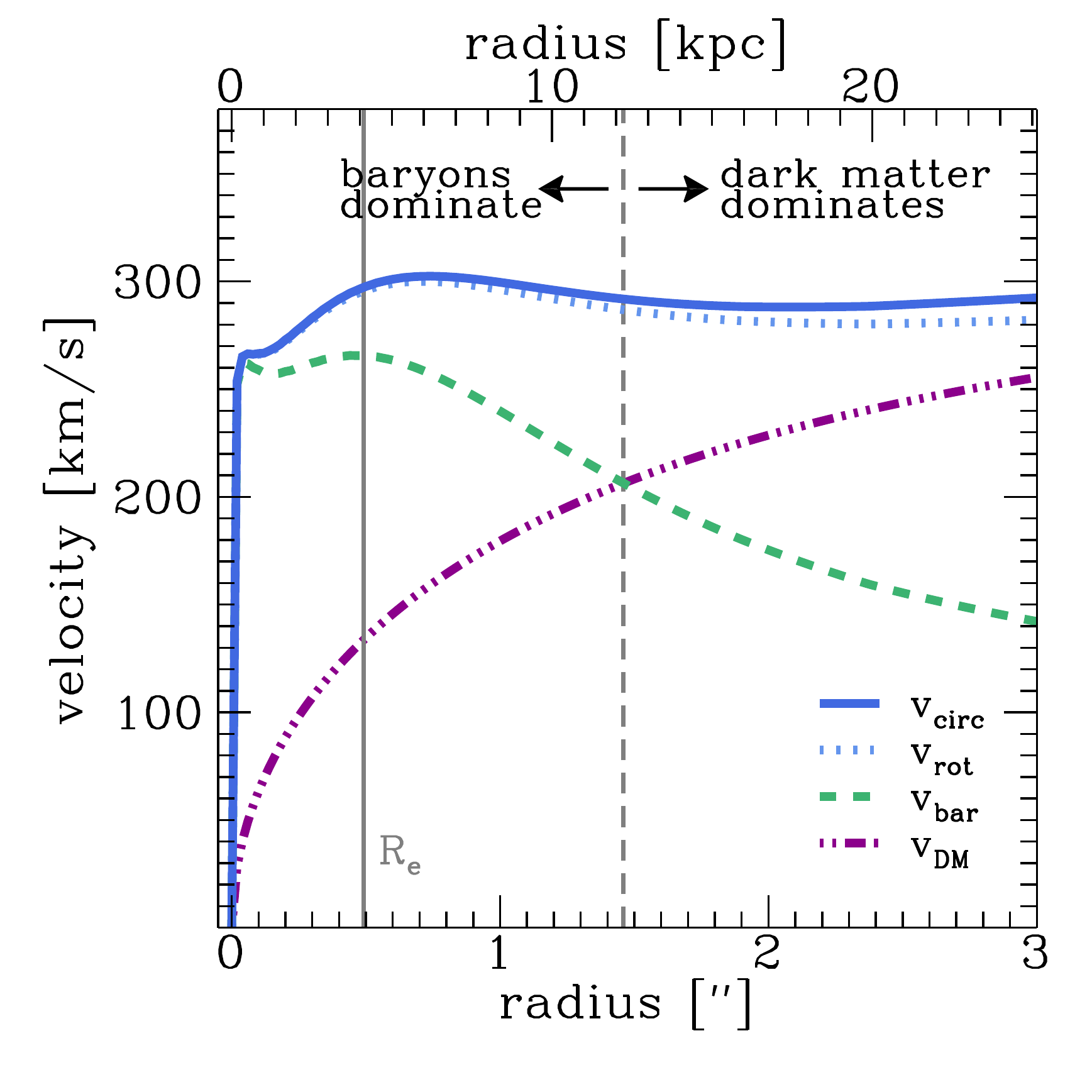}
    \caption{{\bf Left:} Rotation curve in observed {\it vs.\ }intrinsic space. The grey squares show the folded, observed velocity (H$\alpha$+CO) as a function of projected distance from the center. The red dashed line is our fiducial model in observed space. The dash-dotted yellow line shows the model rotation velocity in observed space, corrected for beam-smearing (`bs'). The dotted blue line shows the intrinsic model rotation velocity, further corrected for inclination (`inc'). The solid blue line shows the intrinsic model circular velocity, further corrected for pressure support (`ps'), and the shaded area shows the $1\sigma$ uncertainties of the inclination correction. {\bf Right:} Intrinsic rotation curve of the fiducial model. The solid and dotted blue lines are as in the left panel. The baryonic contribution by the bulge and disk is shown as a dashed green line, and the dark matter contribution as a dash-dotted purple line. The inner solid and the outer dashed vertical grey lines respectively show  $R_e$, and the radius where baryons and dark matter contribute equally to the potential.}
    \label{fig:fig4}
\end{figure*}

We measure the enclosed dark matter fraction at $R_e$ from the intrinsic properties of the {\sc dysmal} model defined by the median sampling results, and find $f_{\rm DM}(\leq$$R_{e}$=$0\farcs49)=v_{\rm DM}^2(R_{e})/v_{\rm circ}^2(R_{e})=0.20$. $v_{\rm DM}$ is the contribution to the circular velocity of the dark matter halo, and $v_{\rm circ}$ is the total circular velocity. The galaxy is strongly baryon-dominated within $R_{e}$. This baryon-dominance prevails out to $r=1\farcs46$ ($3R_e$). Our model agrees with the baryonic disk being `maximal', $v_{\rm disk}(R_{\rm max})/v_{\rm circ}(R_{\rm max})=0.90$, where $R_{\rm max}=0\farcs44$ is the radius where the disk velocity reaches its peak value \citep[e.g.][]{vAlbada85}. The intrinsic model rotation curve and mass component curves are shown in Figure~\ref{fig:fig4}.
The inferred baryon-to-total mass fraction $m_d=0.03$ is compatible with predictions from abundance matching estimates that account for gas mass \citep{Burkert16}.

Through the MCMC sampling, we also gather information on the probability distribution of $f_{\rm DM}(\leq$$R_{e})$, which is not itself a model parameter but calculated from the intrinsic models. In Figure~\ref{fig:fig3} (bottom row) we show the 1D and 2D histograms of the marginalized posterior distribution of the $f_{\rm DM}(\leq$$R_{e})$ values associated with the sampled parameter space. 
While correlations with some of the model parameters are evident, particularly with $M_{\rm bar}$ and with the structural parameters $R_e$ and $B/T$, dark matter is sub-dominant within $R_e$ for the explored parameter space. 
We use the median and 1$\sigma$ confidence ranges of the marginalized probability distribution to estimate $f_{\rm DM}(\leq$$R_{e})$ and its uncertainties, and find $f_{\rm DM}(\leq$$R_{e})=0.18^{+0.06}_{-0.04}$.\\

\section{Discussion \& Conclusions}\label{conclusions}

We have presented kinematic data of a $z=1.4$ SFG based on independent and deep H$\alpha$ and CO(3-2) observations. We find that the ionized and molecular gas trace the same gravitational potential, as their kinematics agree within the uncertainties. Thus, we combine them to model the galaxy.

We use MCMC sampling to constrain a mass model consisting of a thick exponential disk, a bulge, and an NFW halo. We find that the galaxy's central region is baryon-dominated with a dark matter fraction of $f_{\rm DM}(\leq$$R_{e})=0.18^{+0.06}_{-0.04}$. 
This is in agreement with recent findings of low central dark matter fractions in high-redshift SFGs by several groups \citep{FS09, vDokkum15, Alcorn16, Price16, Stott16, WuytsS16, Genzel17, Lang17}.

Together with $v_{\rm{circ}}(R_e)=296$kms$^{-1}$, this places EGS4-24985 into the same region of the $v_{\rm circ}$-$f_{\rm{DM}}$ parameter space as the two $z\sim1.5-1.6$ galaxies observed in H$\alpha$ by \cite{Genzel17} -- a region also populated by massive local SFGs \citep[e.g.][]{Persic88, Begeman91, deBlok08, Lelli16b} and early-type galaxies \citep[e.g.][]{Cappellari13}. The latter are the likely descendants of massive SFGs at $z\sim1-3$. Our result supports the interpretation by \cite{Genzel17} that the low central dark matter fractions observed during the peak epoch of cosmic star formation rate density might be preserved over the rest of cosmic history, as massive SFGs quench and evolve into passive galaxies. Also, this suggests that massive disks are baryon-dominated in their centers at all times.

The low pressure support in our galaxy results in a flat intrinsic rotation curve despite the low $f_{\rm DM}(\leq$$R_{e})$, thus setting it apart from the galaxies presented by \cite{Genzel17}. 
It also implies that in this case the slope of the rotation curve in the outer disk region is a closer tracer of the relative contributions of baryons and dark matter to the rotational support of the galaxy.
The low intrinsic dispersion further suggests that the galaxy is more settled than other galaxies at this redshift with otherwise comparable physical properties \citep{Genzel17}, indicating that any potential dissipative condensation has happened at earlier times \citep[e.g.][]{Dekel14}. 
Still, EGS4-24985 falls on the high-redshift Tully-Fisher relations \citep{Uebler17}.

The agreement of the deep H$\alpha$ and CO data especially in the outer disk helps to alleviate concerns that ionized gas kinematics at high redshift might be unrepresentative of the galaxy kinematics, and could instead be circum-galactic or in-/outflowing gas in disguise. 
Future studies with high-quality resolved kinematics traced through multiple gas phases in SFGs at similar redshifts will be important to statistically corroborate our result.\\


\acknowledgements 
We are grateful to the anonymous referee for a constructive report that helped to improve this manuscript. 
We thank the staff at IRAM and LBT for their helpful support with the NOEMA and LUCI observations for this work.



\bibliographystyle{aasjournal.bst}
\bibliography{literature}



\end{document}